\documentclass[apjl]{emulateapj}

\journalinfo{To appear in the {\em GALEX} Special Edition of ApJ Letters}

\shorttitle{UV Morphology of NGC 4038/39}
\shortauthors{Hibbard et al.}

\begin{document}

\title{UV Morphology and Star Formation in the Tidal Tails of NGC 4038/39}

\author{J.E. Hibbard\altaffilmark{1}, 
Luciana Bianchi\altaffilmark{2}, 
David A. Thilker\altaffilmark{2}, 
R. Michael Rich\altaffilmark{3}, 
David Schiminovich\altaffilmark{4,5}, 
C.Kevin Xu\altaffilmark{4}, 
Susan G. Neff\altaffilmark{6},
Mark Seibert\altaffilmark{4}, 
S. Lauger\altaffilmark{7}, 
D. Burgarella\altaffilmark{7},  
Tom A. Barlow\altaffilmark{4},
Yong-Ik Byun\altaffilmark{8}, 
Jose Donas\altaffilmark{7},
Karl Forster\altaffilmark{4},
Peter G. Friedman\altaffilmark{4},
Timothy M. Heckman\altaffilmark{9},
Patrick N. Jelinsky\altaffilmark{10},
Young-Wook Lee\altaffilmark{8},
Barry F. Madore\altaffilmark{11},
Roger F. Malina\altaffilmark{7},
D. Christopher Martin\altaffilmark{4},
Bruno Milliard\altaffilmark{7},
Patrick Morrissey\altaffilmark{4},
Oswald H. W. Siegmund\altaffilmark{10},
Todd Small\altaffilmark{4},
Alex S. Szalay\altaffilmark{9},
Barry Y. Welsh\altaffilmark{10}, and
Ted K. Wyder\altaffilmark{4}}

\altaffiltext{1}{NRAO, 520 Edgemont Road, Charlottesville VA 22903
{\it jhibbard@nrao.edu}} 
\altaffiltext{2}{Center for Astrophysical Sciences, The Johns Hopkins
University, 3400 N. Charles St., Baltimore, MD 21218}
\altaffiltext{3}{Department of Physics and Astronomy, 
University of California at Los Angeles, Los Angeles, CA 90095}
\altaffiltext{4}{California Institute of Technology, Pasadena, CA91125}
\altaffiltext{5}{Astronomy Department,
Columbia University, 538 W. 120th St, NYC, NY 10025}
\altaffiltext{6}{Laboratory for Astronomy and Solar Physics, 
NASA Goddard Space Flight Center, Greenbelt, MD 20771}
\altaffiltext{7}{Laboratoire d'Astrophysique de Marseille, BP 8, Traverse
du Siphon, 13376 Marseille Cedex 12, France}
\altaffiltext{8}{Center for Space Astrophysics, Yonsei University, Seoul
120-749, Korea}
\altaffiltext{9}{Department of Physics and Astronomy, The Johns Hopkins
University, Homewood Campus, Baltimore, MD 21218}
\altaffiltext{10}{Space Sciences Laboratory, University of California at
Berkeley, 601 Campbell Hall, Berkeley, CA 94720}
\altaffiltext{11}{Observatories of the Carnegie Institution of Washington,
813 Santa Barbara St., Pasadena, CA 91101}

\begin{abstract}
We present {\tt GALEX} $FUV$ (1530 \AA) and $NUV$ (2310 \AA)
observations of the archetypal merging system NGC 4038/39, ``The
Antennae". Both tails are relatively bright in the $UV$, especially in
the vicinity of the Tidal Dwarf Galaxy candidates at the end of the
southern tail. The $UV$ light generally falls within the optically
delineated tails, although the $UV$ light is considerably more
structured, with a remarkably similar morphology to the tidal
\ion{H}{1}. The $UV$ colors suggest that there has been continuing
star formation within the tidal tails, even outside the previously
studied Tidal Dwarf regions. Within the inner disk regions, there are
interesting $UV$ features which appear to be related to the 
extended soft X-ray loops and halo recently discovered by {\tt Chandra}.
\end{abstract}


\section{Introduction}

It has long been known that tidal filaments have blue optical colors,
with regions as blue or bluer than the spiral arms of disk galaxies 
(Zwicky 1956; Arp 1966; Schweizer 1978; Schombert, Wallin, \&
Struck-Marcell 1990; Weilbacher et al. 2000).  What is not known is
whether the blue colors of the tails reflect the fact that they are
comprised of formerly star-forming disk material that was ejected from
the host spirals several $10^8$ yrs ago, or if they reflect on-going
and pervasive tidal star-formation.

Localized regions of current tidal star formation have been
unambiguously identified via the optical emission lines of the
\ion{H}{2} regions ionized by young
($<$10 Myr) stars (e.g. Stockton 1974a,b; Schweizer 1978;
Hibbard \& van Gorkom 1996; 
Iglesias-P\' aramo \& V\' ilchez 2001; Weilbacher, Duc, \& 
Fritze-v.Alvensleben 2003).
And recently individual young stars (Saviane, Hibbard
\& Rich 2004) and star clusters (Knierman et al. 2003; Tran et al. 2003; 
de Grijs et al. 2003) have been discovered within tidal tails. But in
general these studies are only sensitive to the youngest objects or
only target small regions of the tails. It is thus not clear how
prevalent star formation within the tails is, or if star/cluster
formation only occurs at specific sites or at a specific evolutionary
stage of tidal development.

At a newly revised TRGB distance of 13.8 Mpc (Saviane et al. 2004;
used throughout), the Antennae is the nearest on-going major merger,
and an excellent target for studying tidal star formation. Many
studies have concentrated on the spectacular inner regions of this
system (e.g. 
Zhang, Fall \& Whitmore 2001; Kassin et al. 2003). However, we defer a
discussion of those regions to a future paper.  Here we concentrate on
the tidal regions.  We are particularly interested in whether there
are widespread stellar populations with ages less than the dynamical
time of the tidal tails, which is $\sim$300 Myr as derived either from
the maximum tidal extent divided by the maximum tidal velocity (45
kpc/150 km s$^{-1}$) or the numerical model of Barnes (1988; 320 Myr
when scaled to the distance of Saviane et al. 2004).

Schweizer (1978) discovered several \ion{H}{2} regions near the end of
the southern tail of NGC 4038. Both he and Mirabel et al. (1992)
studied this region and suggested two possible sites of forming or
formed ``Tidal Dwarf Galaxies'' (TDGs): dwarf galaxy-sized
self-gravitating objects assembled from tidal debris (e.g. Duc \&
Mirabel 1999). In the following, we refer to these regions as TDG[S78]
and TDG[MDL92], and indicate their locations in Figure 1 (see also
Hibbard et al. 2001).  The stellar population of the TDG[MDL92] region
was studied with the {\tt HST WFPC2} by Saviane et al. (2004), who
find a population of eight young stellar associations (ages less than
30 Myr), as well as a more extended population of intermediate age
(80--100 Myr) bright red stars. Knierman et al. (2003) used the {\tt
HST WFPC2} to target a region midway along the southern tail, finding
no significant population of bright ($M_V<-8$) star clusters.

NGC 4038/39 was targeted as part of the {\tt GALEX} {\it Nearby
Galaxies Survey} (Bianchi et al. 2004a). When combined with existing
ground-based optical colors, the $UV$ colors provide an adequate color
baseline to discriminate between stellar populations with ages ranging
from 10---1000 Myr, spanning the dynamical age of the tails
and allowing us to identify populations formed both before and after
the tails were ejected.  And while the detailed {\tt HST} observations
mentioned above studied the youngest stellar populations at two small
regions within the southern tail, the sensitivity and wide field of
view of the new {\tt GALEX} observations provides a complete census of
on-going and recent star formation along the entire tidal system,
including the previously unstudied northern tail.

\begin{deluxetable}{lrrrr}
\tablecaption{Global Properties\tablenotemark{a} for NGC 4038/39}
\tablewidth{0pt}
\tablehead{
\colhead{Quantity} & \colhead{Disks} & \colhead{South Tail} & 
\colhead{North Tail} & \colhead{TDG}\tablenotemark{b}}
\startdata
$m_{FUV}$ &  12.95$\pm$0.001 & 16.38$\pm0.01$ & 18.89$\pm$0.07 & 17.71$\pm$0.02 \\
$m_{NUV}$ &  12.62$\pm$0.001 & 15.94$\pm$0.02 & 17.66$\pm$0.02 & 17.66$\pm$0.01 \\
$m_{V}$   &  10.39$\pm$0.04  & 12.87$\pm$0.04 & 13.93$\pm$0.04 & 15.55$\pm$0.04 \\
$F-N$ & 0.33$\pm$0.002   & 0.45$\pm$0.02  & 1.24$\pm$0.07  & 0.05$\pm$0.02 \\
$F-V$   & 2.56$\pm$0.04    & 3.52$\pm$0.04  & 4.96$\pm$0.08  & 2.16$\pm$0.04 \\
Log[Age]\tablenotemark{c} & 8.34$\pm$0.08 & 8.49$\pm$0.13  & 8.71$\pm$0.10 & 8.08$\pm$0.90  \\
\enddata
\tablenotetext{a}{colors and magnitudes are on the AB system.}
\tablenotetext{b}{Region encompassing Tidal Dwarf Candidates
TDG[S78] and TDG[MDL92]; see Fig.1a.}
\tablenotetext{c}{Age for single burst model assuming only foreground
Galactic reddening of $E(B-V)$=0.046 (NED). The error includes both 
propagation of photometric errors and the difference in ages derived 
separately from the $FUV-NUV$ and $FUV-V$ colors, but not errors due 
to the uncertainty in the $UV$ color zero point (see
Morrissey et al. 2004).}
\end{deluxetable}

\section{Observations and Results}

The Antennae was observed by {\tt GALEX} (Martin et al. 2004) on
2004-02-22 in four orbits with the total exposure time of 2554
seconds. The data reduction was done using the standard {\tt GALEX}
pipeline (Morrissey et al. 2004). The r.m.s noise in the pipeline data
is 27.83 AB mag arcsec$^{-2}$ in the $FUV$ (1550{\AA}) image and 28.02
AB mag arcsec$^{-2}$ in the $NUV$ (2310{\AA}) image. The FWHM of the
$FUV$ image is $\sim5''$, and that of the $NUV$ image is $\sim6''$.
We combine the {\tt GALEX} observations with the optical and
\ion{H}{1} observations of Hibbard et al. (2001). For this comparison,
the $UV$ data were first edited to remove stars 
and then convolved to the resolution of the
intermediate resolution \ion{H}{1} data (FWHM=$20.7''\times
15.4''$). This means we will be comparing gas and $UV$ properties
averaged over spatial scales of $1.4\times1.0$ kpc$^2$.  The resulting
images reach limits of $\mu_{UV}\sim$29.5 mag arcsec$^{-2}$.

\begin{figure*}
\plotone{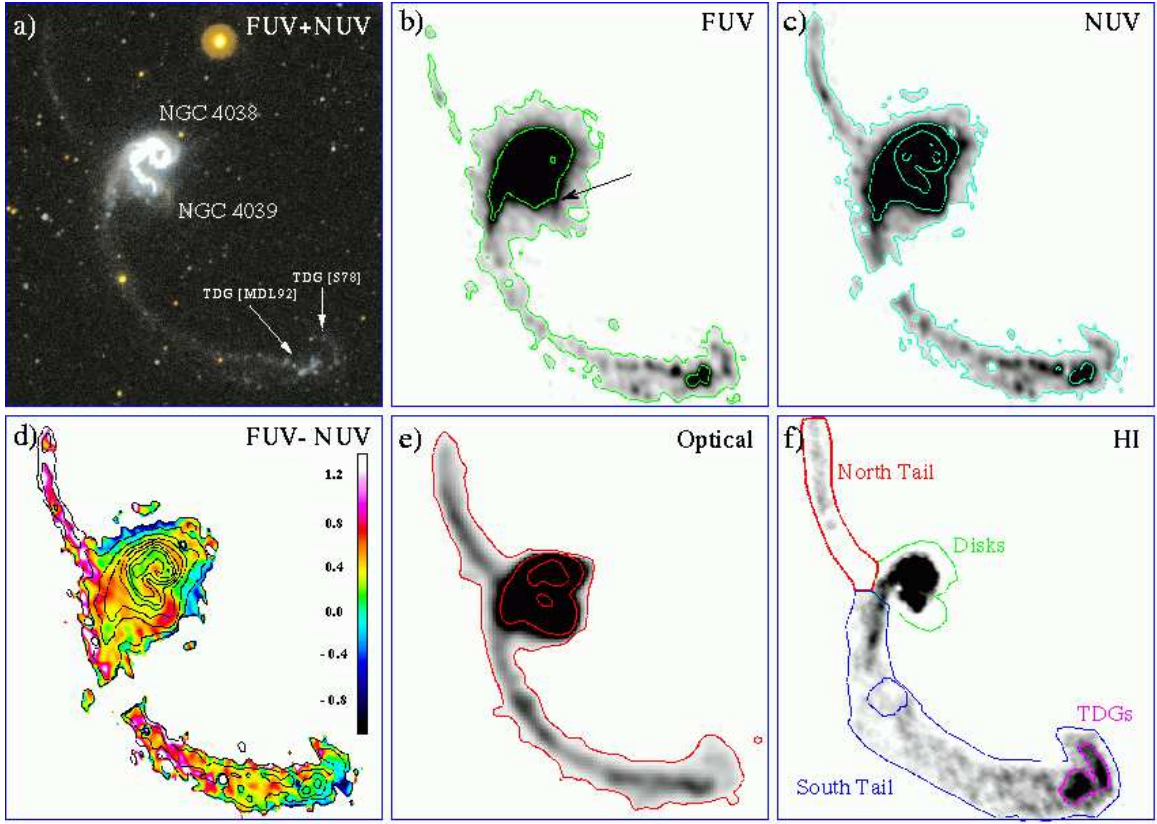}
\caption{Montage of $UV$, optical and \ion{H}{1} observations of 
NGC 4038/39. {\bf (a)} A false-color representation of the full
resolution $FUV$ and $NUV$ data, where blue represents $FUV$ emission,
green a linear combination of $FUV+NUV$ emission, and red $NUV$
emission. The location of the TDG candidates identified by Schweizer 
(1978) and Mirabel et al. (1992) are indicated.
{\bf (b)} Greyscale image of the smoothed $FUV$
emission. Contours are drawn at $\mu_{FUV}$=29.5, 27.0, 24.5 mag
arcsec$^{-2}$. {\bf (c)} Greyscale image of smoothed $NUV$
emission. Contours are drawn $\mu_{NUV}$=29.5, 27.0, 24.5, 22.0 mag
arcsec$^{-2}$. {\bf (d)} A color image of the $FUV-NUV$
colors. Contours of the $NUV$ surface brightness are drawn
$\mu_{FUV}$=29.5 to 21.5 mag arcsec$^{-2}$ in steps of 1 mag
arcsec$^{-2}$. {\bf (e)} smoothed star-subtracted $B$-band image
convolved to 25$''$ resolution, from Hibbard et al. (2001). Contours
are drawn at $\mu_{B}=$26.5, 24, 21.5 mag arcsec$^{-2}$. 
The sharp northern and western edge of the optical disk in
Fig.\,1e is due to incomplete coverage of these regions by the optical
CCD imaging. {\bf (f)}
Intermediate resolution VLA \ion{H}{1} data from Hibbard et
al. (2001). Contours indicate the regions used to measure the global
properties of the labeled regions reported in Table\,1.}
\end{figure*}

Figure 1 presents a montage comparing the $FUV$ and $NUV$
morphology to the optical and \ion{H}{1} data of Hibbard et
al. (2001). The full-resolution $UV$ data is shown in Fig.\,1a, while
the smoothed data are shown in Fig.\,1b,c.
Figure 2 shows the full resolution $UV$ data of the inner disks and 
the end of the southern tail, with \ion{H}{1} column density contours
superimposed. Table\,1 lists global $UV$ and optical magnitudes and
colors of the different regions outlined in Fig.\,1f.

The smoothed $UV$ emission is quite extended and structured,
especially compared to the smoothed optical image
(Fig.\,1e). Remarkably, the \ion{H}{1} map appears similarly
structured (Fig.\,1f). In fact, there is a close correspondence
between \ion{H}{1} column density peaks and $UV$ emission. Most
notable is the correspondence in the vicinity of the tidal dwarf
candidates labeled in Fig.\,1a, and the region immediate to the east
of these regions (Fig.\,2b). From the VLA \ion{H}{1} channel maps,
Hibbard et al. (2001) noticed that the southern tail has a bifurcated
structure, with two parallel gaseous filaments joining up at the
location of the star forming regions associated with TDG[MDL92]. One
filament coincides with the brighter optical spine of the tail, while
another runs along the southern edge with little or no apparent
optical counterpart ($\mu_B>26$ mag arcsec$^{-2}$). The {\tt GALEX}
observations show a very similar bifurcated morphology which matches
up well with the higher \ion{H}{1} column density contours.
NGC 4038/39 is yet another example of a seemingly general 
{\tt GALEX} result that bright $UV$ regions often correspond 
to \ion{H}{1} column density peaks (e.g., Neff et al. 2004; 
Thilker et al. 2004; Xu et al. 2004).


Luminosity-weighted average ages for the stellar populations of the
features delineated in Fig.\,1f were estimated by comparing the
observed $FUV-NUV$ and $FUV-V$ colors with synthetic colors computed
from reddened Bruzual \& Charlot (2003) models, as described by
Bianchi et al. (2004b). The foreground extinction value from the NED
database, E(B-V)=0.046, was adopted as a lower limit. In view of the
associated \ion{H}{1} gas, the intrinsic colors are likely bluer
(especially for the disk and TDG regions), which would lower all the
ages listed in Table\,1.
The population ages were derived from models of Single Burst star
formation (SSP). In the case of composite population, these ages give
an indication of the age of the most recent burst, which dominates the
$UV$ fluxes.  We also compared the observed colors with models of
Continuous Star Formation (CSP).  Most regions are much redder than
any CSP model, thus excluding this scenario.  More extended model
comparisons considering a variety of star formation scenarios are
deferred to a subsequent paper.

Table\,1 suggests a slightly older characteristic population for the
north tail compared to the south tail. This difference is likely
related to the very different gas contents of the two tails
(Fig.\,1f). The age estimates lie close to the expected dynamical
age of the tail of $\sim$300 Myr (\S1). So, in the absence of internal
reddening, the global tail ages are consistent with little or no star
formation subsequent to tidal ejection, with the exception of the
TDG regions (known to harbor star forming regions).

To get a better idea of whether star formation has proceeded in
regions apart from the TDGs since the tails were ejected, we conducted
photometry on other $UV$-bright complexes along the tails. The
resulting color-coded ages are shown graphically in Fig.\,2c. Again we
have assumed the minimum foreground reddening. In this figure, only
regions with red or magenta colors have population ages as old or
older than the dynamical age of the tails. Most of the $UV$-bright
regions have populations which are significantly younger. This figure
suggests that, while the general stellar population of the tails is of
order its dynamical age, there are a number of locations {\it besides
the TDG regions} which have formed stars subsequent to the ejection of
the tails. Fig.\,2c thus provides a valuable finding chart for future
pointed observations in order to study extra-disk star formation. We
conclude that there has been some low-level {\it in situ} star
formation within the tidal tails, mostly associated with the brighter
$UV$ regions and local gas density peaks.

Fig.\,2c further shows an indication of an age gradient, from
$\approx$ 300 Myrs to less than 100 Myrs, when moving outward along
both tails.  This gradient is hinted at by the $FUV-NUV$ color map
(Fig.\,1d) and the $B-R$ color map of Hibbard et al. (2001), both of
which become bluer with distance. For the two brightest clumps in the
`TDG' region, somewhat discrepant ages are obtained from the $FUV-NUV$
and $FUV-V$ colors, suggesting a composite population.  This is
consistent with the results of Saviane et al. (2004), who find an
extended population of intermediate age (80--100 Myr) bright red
stars in addition to the young stars and stellar associations in the
TDG region (ages $\approx 4-30$ Myr).
The presence of similar mixed populations of young, intermediate age,
and older stars was inferred from {\tt GALEX} measurements of tidal
tails in other interacting galaxies (Neff et al. 2004).

\begin{figure*}[t!]
\plotone{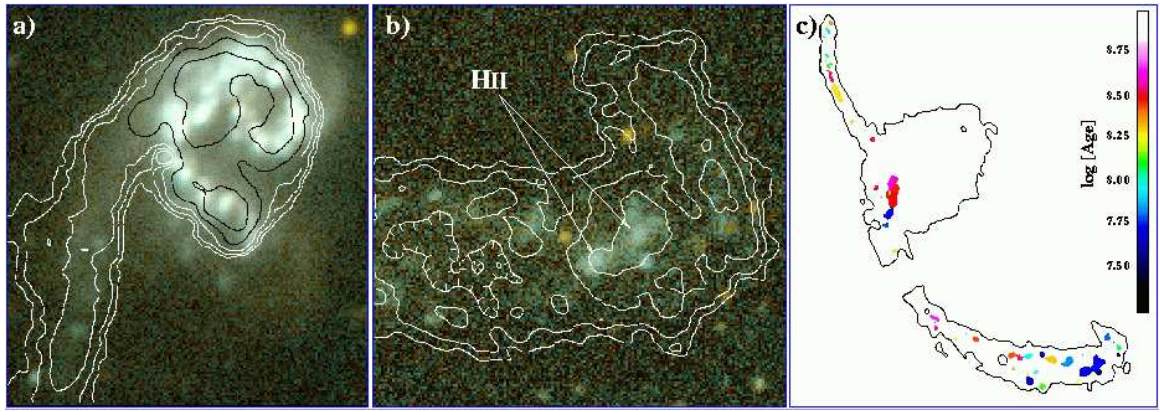}
\caption{(a) \& (b): Detail of inner disks (left)
and TDG regions (middle).
In this false-color image blue represents $FUV$ emission,
green a linear combination of $FUV+NUV$ emission, and red $NUV$
emission. \ion{H}{1} contours are
drawn at (1, 2, 4, 8, 16) $M_\odot {\rm pc}^{-2}$. 
The location of the \ion{H}{2} regions discovered by 
Schweizer (1978) are indicated in the middle panel.
(c) SSP Ages for selected $UV$ bright regions of the
north and south tails, assuming only foreground extinction. The
contour is drawn at $\mu_{NUV}=$27.5 mag arcsec$^{-2}$.}
\end{figure*}

Finally, the inner regions of the progenitor disks are shown in
Fig.\,2a. While these regions will be the subject of a future paper,
we note here several features brought out in the smoothed $UV$ data:
the first is that the $FUV$ flux is sharply truncated in the
southwestern part of the disk of NGC 4038 (see arrow in Fig.\,1b), A
similar sharp edge is seen in the \ion{H}{1} disk (Fig.\,2a). The $UV$
color of this region is significantly redder than the rest of the disk
($FUV-NUV\sim$0.9 mag, compared to 0.3 mag). Imaging of this region at
a variety of wavelengths (e.g. Zhang, Fall \& Whitmore 2001) suggest
that this is not due to dust, and the optical-NIR imaging of Kassin et
al. (2003) suggest that the stellar population of this region is
uniformly very old. This region is associated with the giant loops of
hot thermal plasma ($kT\sim0.3$ keV) imaged by {\tt Chandra} (Fabbiano
et al. 2004), raising the possibility that an X-ray wind swept this
region clear of cold gas in the recent past, truncating star
formation. On a larger scale, there is a halo of faint $UV$ emission
surrounding both disks (Fig.\,1b,c). This emission is unlike that in
the tails in that it is more extended in the $FUV$ than the $NUV$
(note the blue colors in Fig.\,1d). It is uncertain whether there is
faint optical counterpart to this emission, since the optical CCD
imaging did not extend this far.  If there is no optical counterpart,
then the emission may be associated with shocked gas in the recently
discovered X-ray halo (Fabbiano et al. 2004).

\section{Conclusions}

We have studied the $UV$ light within the archetypal merger NGC
4038/39, ``The Antennae''.  The tidal $UV$ morphology is remarkably
similar to the tidal \ion{H}{1}, with a close correspondence between
regions of bright $UV$ emission and high
\ion{H}{1} column density. There are interesting features in the 
$UV$ morphology of the inner regions. The $FUV$ light in the
southwestern half of the disk of NGC 4039 is sharply truncated,
coincident with a similar truncation in the
\ion{H}{1} disk. This gas may have been removed by the X-ray loops
recently imaged in this system by {\tt Chandra} (Fabbiano et
al. 2004), inhibiting subsequent star formation. On a larger scale,
there is a $FUV$ ``halo'', possibly related to shocked gas in the
X-ray halo.

While it has long been known that tails are generally blue (\S1), the
{\tt GALEX} observation provide color baselines to determine how much
of the $UV$ radiation comes from stars younger than the dynamical age
of the tails. A preliminary analysis suggests that most of the stars
within the tidal tails date to before the tail formation period. The
population in the northern tail appears older than that in the
southern tail. Examining individual $UV$-bright regions within the
tails we find regions of more recent star formation, which occur 
at regions of higher \ion{H}{1} column density, and extend
well beyond the previously identified tail star forming regions. The
tails are thus a promising laboratory for studying extra-disk star 
formation and testing theories of star and cluster formation.

The present results are for a very simplistic star formation
history. A future analysis will include more complicated star
formation histories, and will incorporate all available
$UV$-to-optical colors of the tails. The $UV$ color baselines should
enable us to not only identify current star forming regions, but to
examine and date regions of past star formation. By comparing the
number of regions of a given age and the star formation history in
individual knots, we will be able to test whether long lived
condensations form within tidal tails. 

\acknowledgements

JEH thanks Fran\c cois Schweizer for useful discussions, and
the referee for a useful and timely report. {\tt GALEX}
(Galaxy Evolution Explorer) is a NASA Small Explorer, launched in
April 2003.  We gratefully acknowledge NASA's support for
construction, operation, and science analysis for the {\tt GALEX}
mission, developed in cooperation with the Centre National d'Etudes
Spatiales of France and the Korean Ministry of Science and Technology.


\end{document}